# Single crystal field-effect transistors based on an organic selenium-containing semiconductor

R. Zeis[*], Ch. Kloc[*], K. Takimiya[+], Y. Kunugi[♦], Y. Konda[+], N. Niihara[+] and T. Otsubo[+]

.

[*]Bell Laboratories, Lucent Technologies, 600 Mountain Ave., Murray Hill, NJ 07974, USA,

[+]Graduate School of Engineering, Hiroshima University, 1-4-1 Kagamiyama, Higashi-Hiroshima 739-8527, Japan,

[♦]Faculty of Integrated Arts and Sciences, Hiroshima University, 1-7-1 Kagamiyama, Higashi-Hiroshima 739-8521, Japan

**We report on the fabrication and characterization of single crystal field-effect transistors (FETs) based on 2,6-diphenylbenzo[1,2-b:4,5-b']diselenophene (DPh-BDSe). These organic field-effect transistors (OFETs) function as p-channel accumulation-mode devices. At room temperature, for the best devices, the threshold voltage is less than –7V and charge carrier mobility is nearly gate bias independent, ranging from 1cm$^2$/Vs to 1.5 cm$^2$/Vs depending on the source-drain bias. Mobility is increased slightly by cooling below room temperature and decreases below 280 K.**

## 1. Introduction

Organic semiconductors have received considerable attention within recent decades. Already the first “plastic electronic” products, such as displays that use organic light-emitting diodes, are on the market, setting the stage for organic semiconductor technology. This technology has potential applications for inexpensive printed devices like radio-frequency identification tags (RFIDs), low-end and high-volume data storage devices, flexible, cheap and therefore disposable displays and wearable computing. Organic solar cells and chemical sensors have recently attracted special attention.

For these above-mentioned applications, a better understanding of charge transport in organic semiconductor materials is required. Field-effect transistors (FETs) allow continuous tuning of charge density induced by the transverse electric field and enable the systematic study of charge carrier transport in organic materials. For our study, as a semiconductor, we chose 2,6-diphenylbenzo[1,2-b:4,5-b’]diselenophene (DPh-BDSe), with the molecular structure shown in Fig. 1. The synthesis of DPh-BDSe has been reported recently[1].

This material contains large selenium atoms which we hoped would improve the intermolecular interactions and allow efficient pi electron transport between molecules, an idea which had already been demonstrated in thin film field effect transistors[1]. A relatively high mobility of 0.17 $cm^2/Vs$ and an on/off ratio larger than $10^5$ had already been measured[1] but we assumed that even better parameters might be obtained by avoiding grain boundaries and limiting the concentration of impurities and defects. Therefore, we use high quality single crystals for fabrication of field effect transistors.

## 2. Experimental details

The molecules were synthesized from commercially available p-dibromobenzene using a three-step reaction[1]. For structure determination, single crystals were grown from benzonitride solution and herringbone structure was determined using a Rigaku AFC7 single crystal diffractometer. DPh-BDSe single crystals were grown by horizontal physical vapor transport in the flow of argon gas. Details of the growth technique have been reported before[2]. The evaporating material DPh-BDSe was heated to 300ºC in the hot zone of a two-zone furnace. The second zone of the furnace was held at 200ºC. We use two sublimations to purify the starting material. DPh-BDSe single crystals spontaneously grow on the wall of glass tube in the colder zone of the furnace as pale yellow platelets with a surface up to 5*5 $mm^2$ and a thickness of the order of 10μm. The structure of the gas-phase-grown DPh-BDSe single crystals was confirmed by X-ray diffraction to be the same as that of the solvent-grown crystals[1].

On the surface of the fresh grown crystals, a typical[3] field effect transistor structure was produced (Fig. 2). Source and drain contacts were painted with a water-based solution of colloidal graphite. The gate-insulating layer consisted of a 1.1-1.7 μm Parylene N thin film. The thickness of the films was determined with a profilometer. On top of the parylene layer, between source and drain, the gate electrodes were painted with the colloidal graphite. The channel capacitance was calculated from the thickness of the insulating layer and the tabulated dielectric constant of Parylene N. The transistor characteristic was measured at room temperature using a test fixture connected to a HP 4145B semiconductor parameter analyzer. The low temperature measurements were performed in a helium atmosphere in a Quantum Design cryostat, which was also connected to a HP 4145B semiconductor parameter analyzer. To precisely control the temperature of the devices inside the Quantum Design cryostat, we used an additional temperature sensor placed near the sample.

## 3. Results and Discussion

Fig. 3 shows the source-drain current ($I_{SD}$) as a function of the applied source-drain voltage ($V_{SD}$) for different gate voltages ($V_G$). For small negative source-drain voltages ($V_{SD}$) the FET operates in the linear regime. When the source-drain voltage increases, the gate field is no longer uniform and a depletion area is formed at the drain contact. Beyond a certain source-drain voltage ($V_{SD}$), the current becomes saturated. This behavior is typical for a p-type field-effect transistor.

Plotted in Fig. 4 are the trans-conductance characteristics, for different source-drain voltages ($V_{SD}$). As these characteristics indicate, the DPh-BDSe single crystal device does not develop a sharp field-effect onset. For small gate voltages, the source-drain current changes only gradually with the applied field. This feature might indicate the existence of a resistivity barrier on the contacts. The high contact resistance in an Schottky-type OFET depends non-linearly on the gate voltages. A similar effect that dominates especially the subtheshold region has also been observed for rubrene single crystal FETs[4]. However, the DPh-BDSe single crystal devices exhibit, at $V_{SD}$=-5V, a subthreshold swing (S) of around 7V/decade, which corresponds to a normalized subthreshold swing ($S_i$) of 11 V·nF/decade·cm$^{-2}$. This value is comparable with α-Si:H FETs and CuPc single crystal FETs, for which $S_i$=10 V·nF/decade·cm$^{-2}$ [5] and $S_i$=7 V·nF/decade·cm$^{-2}$ [6]respectively have been reported. For comparison, in the best devices on Rubrene single crystal FETs, which usually develop a sharper field-effect onset, the normalised subthreshold swing is $S_i$=1.7 V·nF/decade·cm$^{-2}$ [7].

The field-effect onset is observed at a negative gate voltage (–1V). For a p-type device, this behavior resembles a "normally-off" FET, which seems to be the case for all organic single crystal field-effect transistors with parylene as a gate dielectric[6-8]. Assuming that the density of electrical active traps is proportional to the charge needed to fill these traps, and taking the threshold voltage of –7V from Fig. 4, the density of the charged traps at the DPh-BDSe/parylene interface is estimated to be 6.6 *$10^{10}$ cm$^{-2}$. In this rough evaluation the contribution of the contacts on the threshold voltage has not been taken into account, but such a procedure allows us to estimate the upper limit of trap concentration mobility.

Due to the relative high bulk conductivity of this material, the on/off ratio below $10^4$ was obtained from the trans-conductance characteristics. From the data presented in Fig.4, we determined the charge carrier mobility in the linear regime ($|V_{DS}|<|V_G-V_T|$) by using the equation $\mu=(L/WC_iV_{SD})(dI_{SD}/dV_G)$ [9], where $C_i$ is the gate insulator capacitance per unit area. As shown in Fig. 5, the field-effect mobility depends on the source-drain voltages. With increasing source-drain bias, the mobility increases until it saturates at 1.5 cm$^2$/Vs. This means that for a sufficiently large longitudinal electric field the transistor performance is no longer limited due to the Schottky barrier of the contacts. If we calculate the mobility from the saturation regime ($|V_{DS}|>|V_G-V_T|$) applying the equation $\mu=2(L/WC_i)(dI_{SD}/dV_G)^2$ [9], the equivalent behavior was found. In addition, we obtained the same maximum carrier mobility of 1.5 cm$^2$/Vs. V. Podzorov et. al.[7] and C. Goldman[10] et. al. have reported a similar source-drain bias dependence for rubrene single crystal

FETs. Fig. 5 also indicates that for a sufficiently large negative gate bias ($V_G$<-20V) the carrier mobility becomes nearly independent of the $V_G$. This feature reflects the high quality of the crystal, with only few structural defects[11].

It is worth noting that a mobility of 1.5 $cm^2/Vs$ is the highest value we obtain for our DPh-BDSe single crystal devices, but that mobilities above 1$cm^2/Vs$ were routinely measured. Furthermore, compared to thin film devices, the mobility for single crystal FETs is about one order of magnitude higher[1].

Fig. 6 shows the slight increase of the carrier mobility with cooling until it reaches its maximum at 280K. Below 280K, the mobility drops rapidly and displays a thermally activated behavior. For a large number of trapping states present in the band gap, an Arrhenius-like dependence of the mobility $\mu \sim \exp(-E_A/k_bT)$ is to be expected. We determined an activation energy ($E_A$) of around 25meV. However, we would like to add that, at low temperatures the Schottky barrier on the contacts may limit the source drain current and therefore the activation energy calculated from the two electrodes; mobility may need to be corrected for this effect.

**Conclusions**

To increase the carrier mobility in aromatic compounds, we have introduced selenium atoms into the aromatic rings. We grew single crystals and prepared single crystal FET. We have found that a mobility as high as 1.5 $cm^2/Vs$ can be achieved and that mobility slightly increases with cooling following a strong decrease at low temperatures.

## Acknowledgments

We thank Prof. E. Bucher for his support and advice, W. So for fruitful discussions and C. G. Maclennan for reading the manuscript. R. Zeis acknowledges the financial support from the Konrad Adenauer Foundation, the German Academic Exchange Service (DAAD) and the Landesgraduiertenfoerderung Baden-Wuertemberg. We acknowledge the support of the US Department of Energy under grant # 04SCPE389.

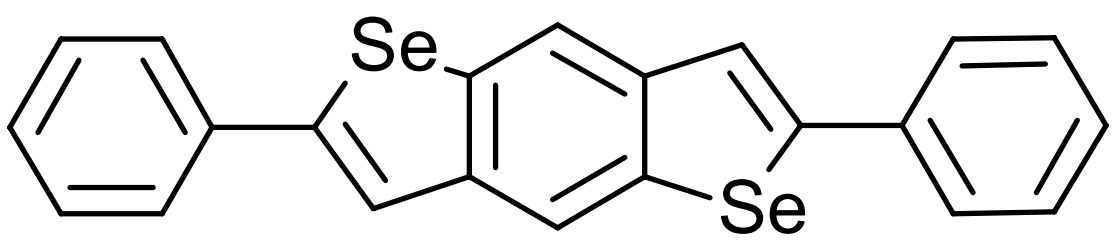

**Fig. 1.** The molecular structure of DPh-BDSe

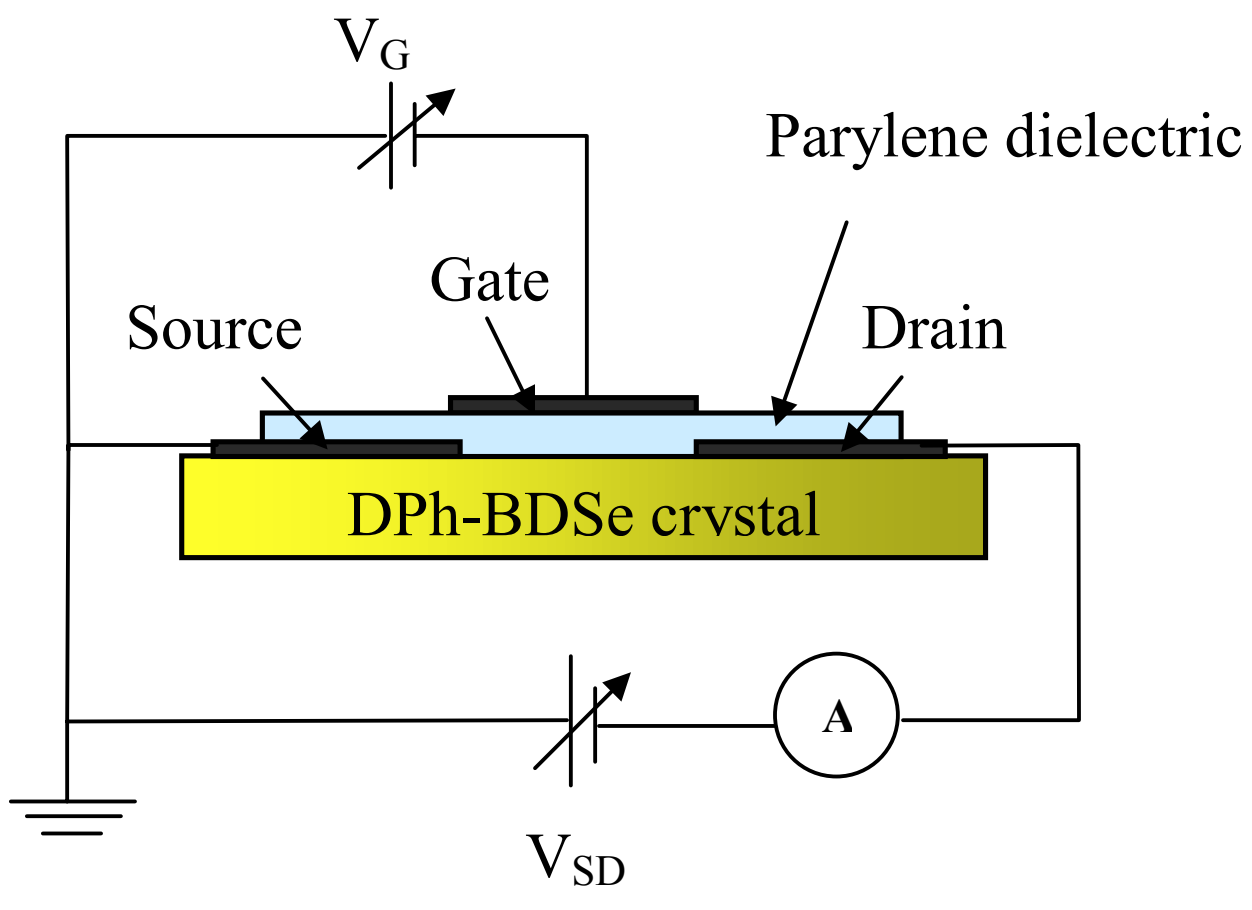

**Fig. 2.** The schematic representation of a DPh-BDSe-based FET and the measuring circuit.

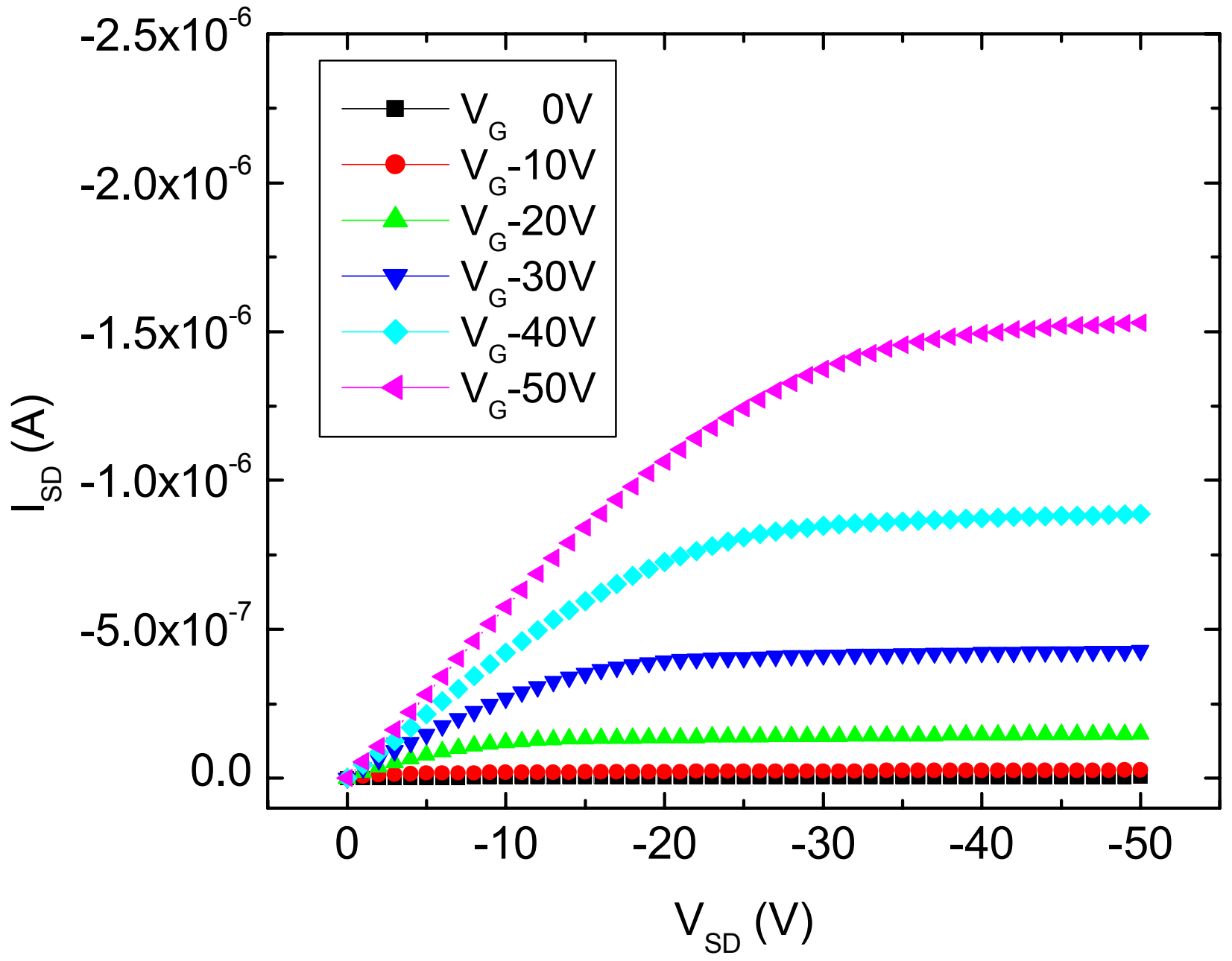

**Fig. 3.** The output-characteristic of a DPh-BDSe FET.

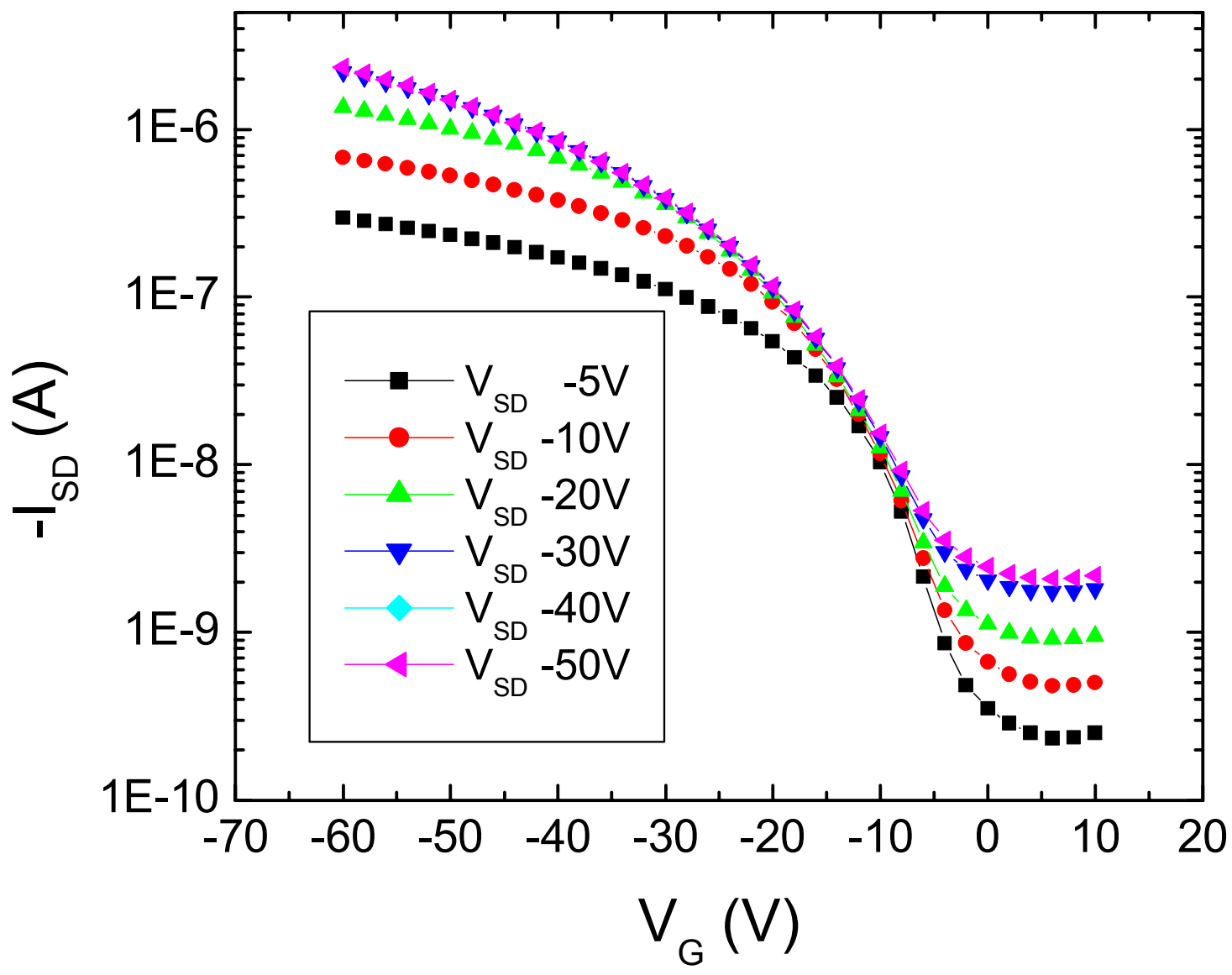

**Fig. 4.** The trans-conductance characteristics of a DPh-BDSe FET (same devices) , for different source-drain voltages.

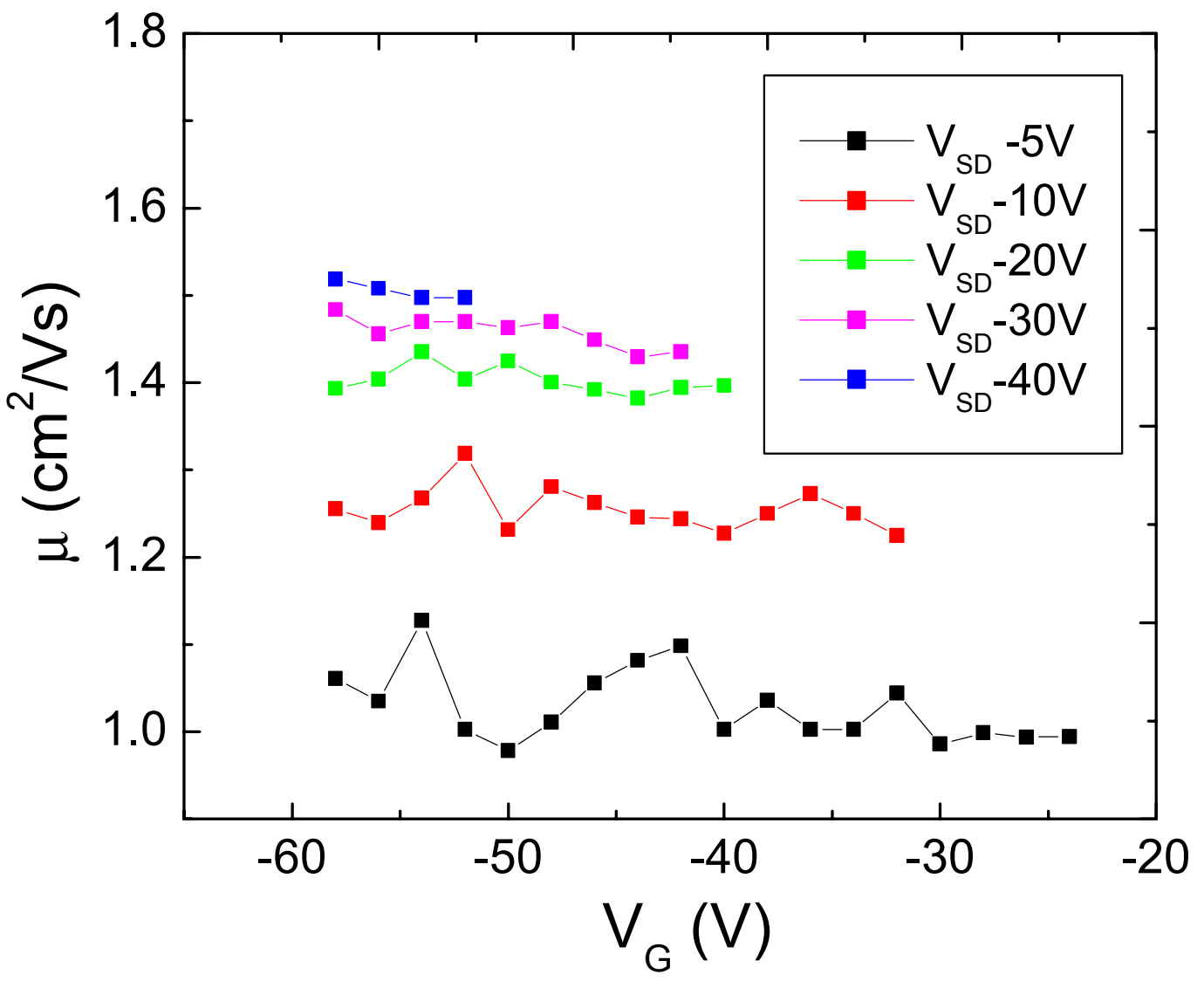

**Fig. 5.** The charier mobility of a DPh-BDSe FET in linear regime (obtained from Data in Fig. 4)

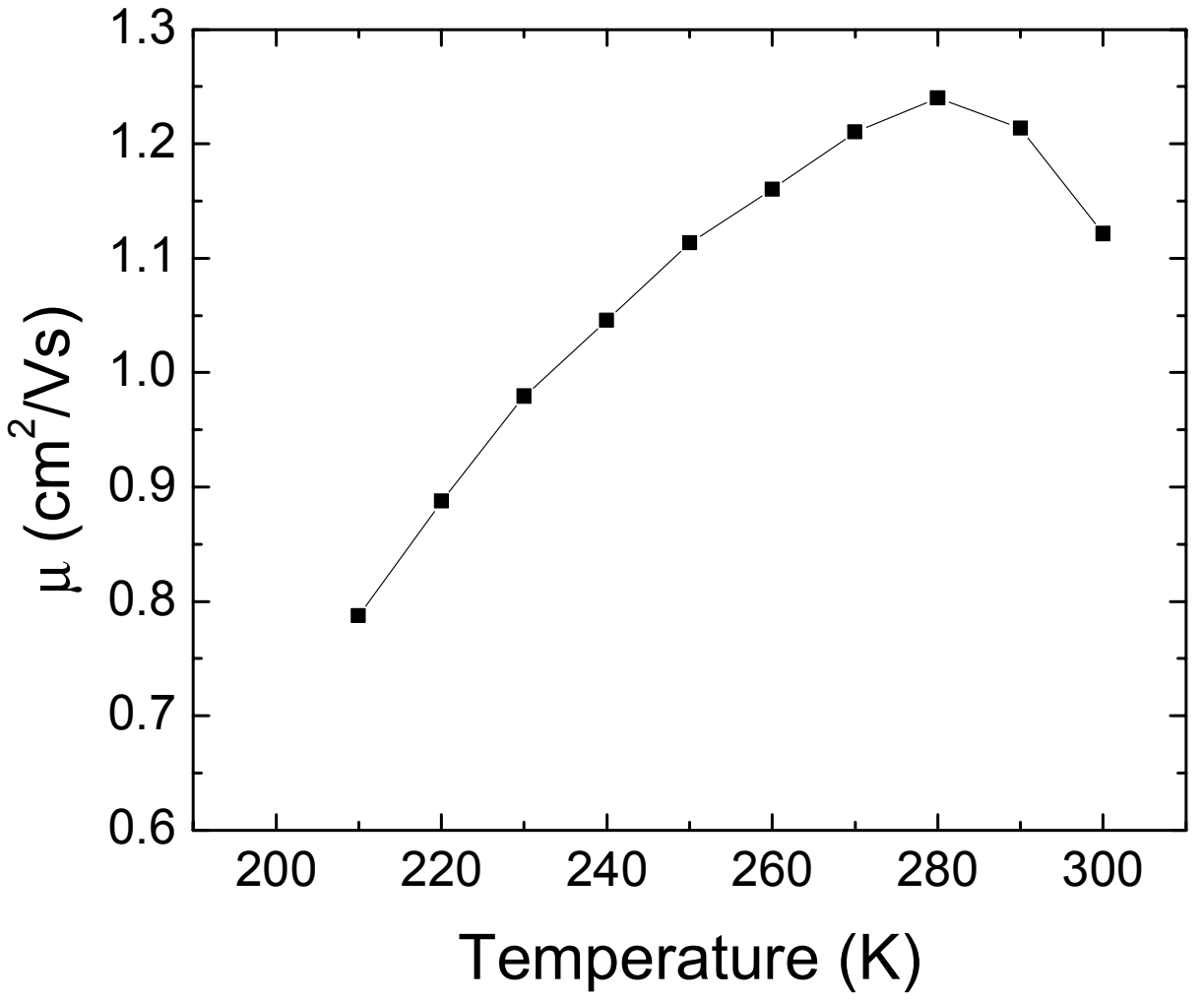

**Fig. 6.** The temperature dependence of carrier mobility calculated from the square root of the saturation current ($V_{SD}$= -60V)